\documentclass{cjaa}                    

\usepackage{graphicx}                   
\input{epsf.sty}                        
\input{psfig.sty}                       

\begin{document}

   \title{Supernovae in Three-Dimension 
}
   \subtitle{A Link to Gamma-Ray Bursts}

   \author{Keiichi Maeda
      \mailto{}
      }
   \offprints{Keiichi Maeda}                   

   \institute{Max-Planck-Institut f\"ur Astrophysik, 
Karl-Schwarzschild-Stra{\ss}e 1, 85741 Garching, Germany
             \email{maeda@MPA-Garching.MPG.DE}
          }

   \date{}

   \abstract{
Observational consequences of a jet-driven supernova (SN) explosion model 
are presented. The results are compared in detail with optical observations of 
SN 1998bw associated with a Gamma-Ray Burst. 
It is shown that the jet model is able to reproduce virtually all the 
optical observations available for this SN, although a spherical model 
fails to explain some of observed features. Because of the viewing angle effect, 
the required kinetic energy of the SN ejecta 
is reduced to $\sim 2 \times 10^{52}$ erg as compared to that obtained by the 
previous spherical model ($\sim 5 \times 10^{52}$ erg), but this is still much 
larger than that of a canonical SN ($\sim 10^{51}$ erg). 
   \keywords{supernovae: individual (SN 1998bw) 
--- gamma rays: bursts 
--- radiative transfer 
--- nuclear reactions, nucleosynthesis, abundances }
   }

   \authorrunning{K. Maeda}            
   \titlerunning{SNe and GRBs in 3-Dimension}  


   \maketitle
%
%
\section{Introduction}           
\label{sect:intro}

Supernovae (SNe) are discovered to date 
in association with three Gamma-Ray Bursts (GRBs), 
i.e., GRB980425 (SN Ic 1998bw), 
GRB 030329 (SN Ic 2003dh), and 
GRB031203 (SN Ic 2003lw) (see e.g., Woosley \& Bloom 2006 for a review). 
An SN Ic (see e.g., Filippenko 1997 for a review of SN terminology) 
is a class of SNe resulting from a core-collapse of 
a C+O star, which has lost 
its hydrogen envelope (and probably even a large fraction of the 
helium envelope) before the explosion 
(e.g., Nomoto et al. 1995). 

The three SNe Ic associated with GRBs 
share similar properties in the early phase. 
The spectra show very broad absorption features, 
and the light curve width is also broad as compared 
to canonical SNe Ic (e.g., Mazzali et al. 2006). 
These properties are explained by a combination of 
large kinetic energy ($E_{\rm K}$) 
and large mass of the SN ejecta ($M_{\rm ej}$), i.e., 
$E_{51} \equiv E_{\rm K}/10^{51}$ erg $\sim (3 - 5)$ 
and $M_{\rm ej} 
\sim 10 M_{\odot}$ (Iwamoto et al. 1998, 
Mazzali et al. 2003, Mazzali et al. 2006). 
The energy is more than 10 times larger than in canonical SNe, 
thus these energetic SNe are sometimes called "hypernovae". 
$M_{\rm ej} \sim 10 M_{\odot}$ (i.e., the mass of the 
C+O star $M_{\rm CO} \sim 12 - 14 M_{\odot}$) 
corresponds to the main-sequence mass $M_{\rm ms} \sim 30 - 40 M_{\odot}$. 
The peak luminosity indicates that $M$($^{56}$Ni) $\sim 0.3 - 0.6 M_{\odot}$, 
larger than in canonical SNe. 

In this paper, we show observational consequences of 
a jet-driven SN explosion model as suggested by Maeda et al. 
(2002) (see also Maeda \& Nomoto 2003b) for SN 1998bw. 
First we summarize the hydrodynamic model of the jet-driven explosion (\S 2). 
Then, after summarizing numerical methods to compute radiation processes in 
a multi-dimensional SN model, expected observational consequences are shown in \S 3 
with detailed comparisons with observations of SN 1998bw.  
The paper is closed with conclusion in \S 4.

\section{Hydrodynamic Model}
\label{sect:model}

\begin{figure}
   \begin{center}        
   \mbox{\epsfxsize=0.5\textwidth\epsfysize=0.5\textwidth\epsfbox{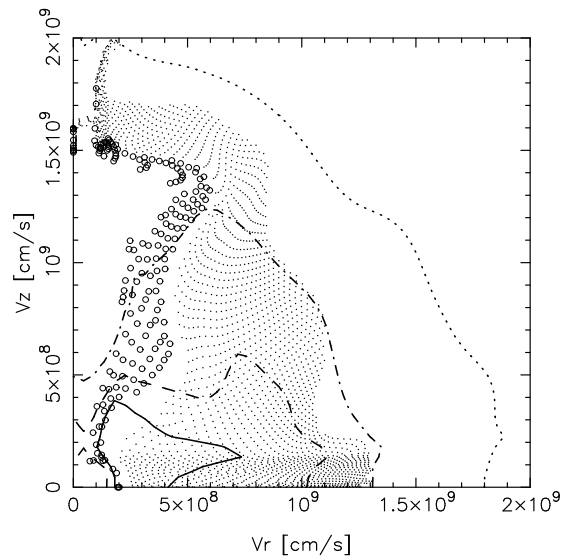}}
   \caption{Structure of the jet-like SN ejecta at a homologous 
expansion phase (valid after $\sim 1$ day and thereafter). 
Distribution of $^{56}$Ni (open circles) 
and of $^{16}$O (dots) are shown. The contours are for density 
distribution. See Maeda et al. (2002) for details.}
   \end{center}
\end{figure}

The massive progenitor for SNe associated with GRBs 
($M_{\rm ms} \sim 40 M_{\odot}$) 
suggests that the central remnant is 
a black hole (BH) rather than a neuron star (NS). 
The large kinetic energy suggests that the explosion mechanism is 
totally different from a canonical SN. 
Most favorable scenario is a formation of a BH 
and a hyper-accreting disk, 
e.g., the collapser model (Woosley 1993, MacFadyen \& Woosley 1999). 

The system of this kind is expected to 
produce an aspherical explosion, irrespective of detailed 
physical processes to create a highly relativistic jet (responsible to a GRB) 
and a sub-relativistic bulk flow (responsible to a hypernova). 
Nucleosynthesis in the stellar mantle as a consequence of 
the jet-driven explosion was studied by several authors 
(Nagataki 2000, Maeda et al. 2002, Maeda \& Nomoto 2003b, Nagataki et al. 2003). 
The general feature is that the explosive nucleosynthesis takes place 
more actively in the jet direction ($z$) than in the equatorial direction ($r$), 
resulting from the stronger shock and the higher temperature in the 
jet direction (Fig. 1). In the $z$-direction, $^{56}$Ni is synthesized abundantly, 
while in the $r$-direction unburned oxygen is left.

\section{Radiation Transfer in Three-Dimension}
\label{sect:radiation}

\begin{figure}
   \begin{center}
   \plottwo{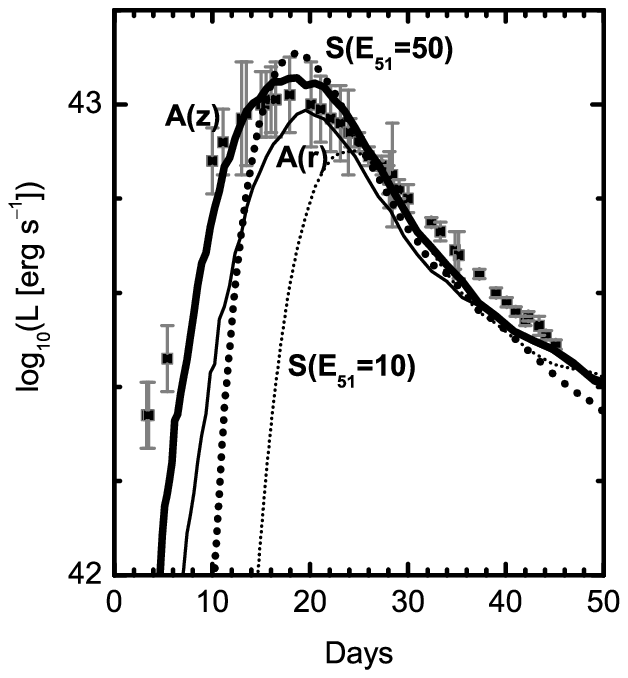}{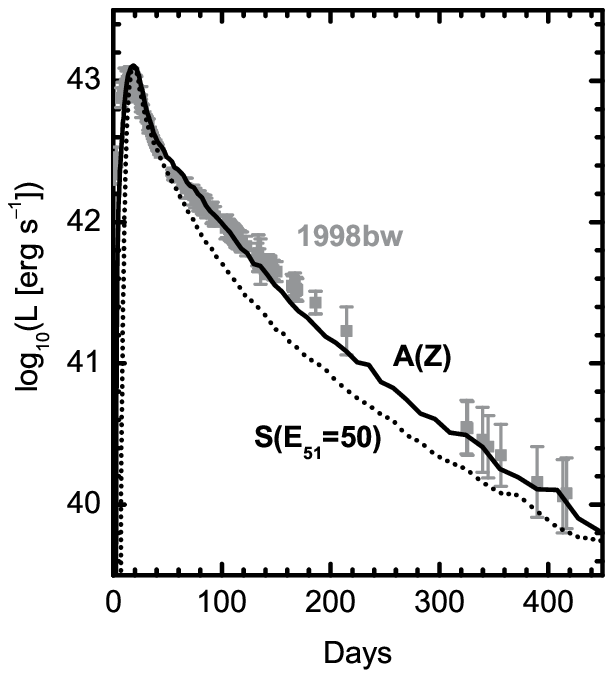} 
   \caption{The light curve of SN 1998bw (gray points) 
in the early (left) and late (right) phases. 
Shown here are the jet-driven model with $E_{51} = 20$ 
as seen at the $z$- [thick black, denoted as A(z)]
and $r$- [thin solid, A(r)] directions. 
Also shown are the spherical models (S). }
   \end{center}
\end{figure}

To test a multi-dimensional explosion model, one has to solve radiation 
transport in multi-dimensional space. 
We have developed the {\it SAMURAI} code 
-- the {\it SupernovA MUlti-dimensional RAdIative transfer} code. 
We adopt the Monte-Carlo method to follow photon packets. 
Transport of $\gamma$-rays (Maeda 2006b) and optical photons (Maeda et al. 2006c) 
is solved to synthesize a bolometric optical light curve and 
multi-energy light curves in the keV -- MeV range. 
Then spectrum calculations are performed for optical photons 
at the early phase (Tanaka et al. 2006) and at the late phase (Maeda et al. 2006a). 
See also H\"oflich et al. (1999), Thomas et al. (2002), Kasen et al. (2004), 
Kozma et al. (2005), and Sim (2006) for other multi-dimensional codes 
available to date. 

Figure 2 shows the optical light curve of SN 1998bw associated with 
GRB980425 (Maeda et al. 2006c). 
The spherical model with $E_{51} = 50$ fits the early phase 
light curve fairly well. 
The aspherical model with $E_{51} = 20$ yields a fit 
as nice as, or even better than, 
the spherical model with $E_{51} = 50$. 
In the early phase, the ejecta are optically thick. 
Thus, the diffusion time scale is determined  
by the optical depth along the line-of-sight. 
Since $E_{\rm K}/M_{\rm ej}^3$ (see Discussion) is effectively large for the jet model 
if viewed on-axis, the diffusion time scale is 
effectively small in this direction. 
Thus the smaller energy is required than the spherical model. 

The LC computation by the {\it SAMURAI} code thus shows 
that the early phase observation can be reproduced either 
by the spherical model or by the aspherical model 
(see also H\"oflich et al. 1999). 
This is found to also be the case for the early phase spectra (Tanaka et al. 2007). 
The degeneracy can be resolved by modeling the late-phase observations. 
Figure 2 shows that the spherical model with $E_{51} = 50$ is 
too faint as compared to the observation, since 
$M_{\rm ej}^2/E_{\rm K}$, which gives the efficiency of 
converting $\gamma$-rays to optical photons, is too small (Maeda et al. 2003a). 
In the late phase,  
$\gamma$-rays can reach at any point of the ejecta not only near $^{56}$Ni. 
The optical photons emitted at every point can reach to the observer. 
The luminosity is thus determined by the global properties 
in the late phases, rather than the local isotropic properties 
as in the early phase. 
In the jet-driven model, 
$M_{\rm ej}^2/E_{\rm K}$ to fit the early phase observations is larger 
than the spherical model. This leads to the large luminosity, 
as large as observed, at the late phase for the jet model (Maeda et al. 2006a). 
It should be emphasized that the early and late phase observations 
provide different information on the SN ejecta -- 
the isotropic values for the former and the intrinsic global values for 
the latter. The effect, as found by our study with the {\it SAMURAI} code, 
is important if the explosion is not spherical. 

\begin{figure}
   \begin{center}
   \plottwo{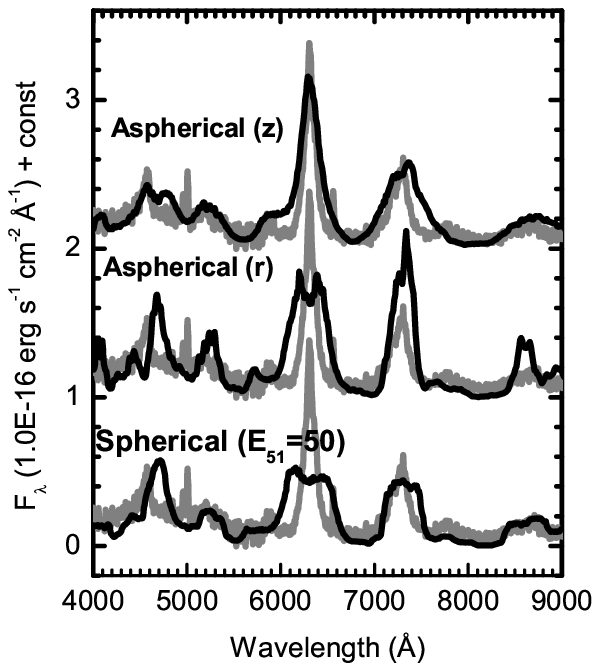}{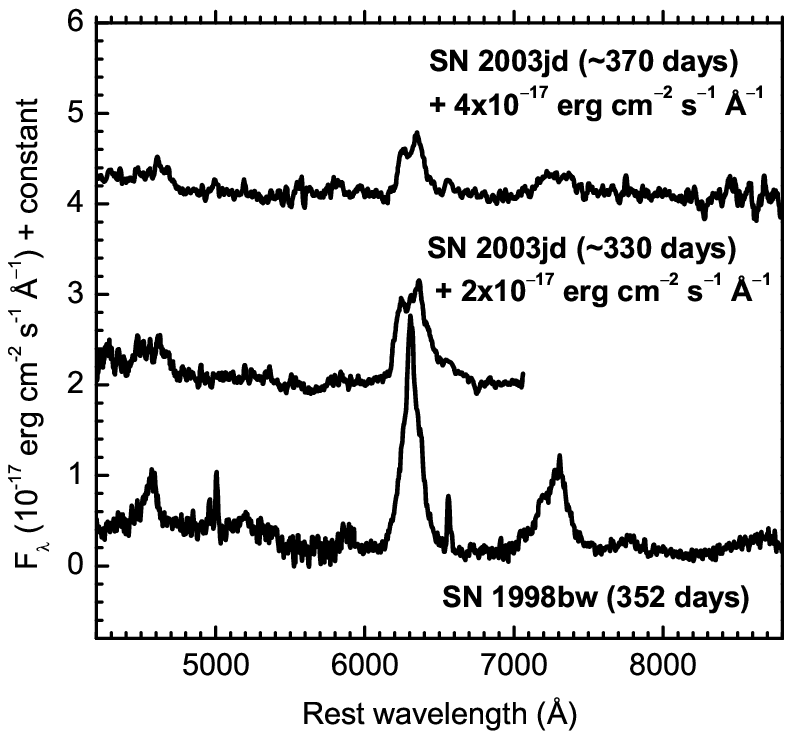}
   \caption{(Left) Late phase spectrum of SN 1998bw at $\sim 1$ years 
after the explosion, as compared with the synthesized spectra of 
the jet-driven model with $E_{51} = 20$ viewed at $z$- and $r$-directions, and 
the spherical model with $E_{51} = 50$ (black curves, from top 
to bottom).
(Right) Late phase nebular spectra of SN 2003jd taken at the Keck 
telescope (top) and at the Subaru telescope (middle), as compared with 
that of SN 1998bw (bottom).}
   \end{center}
\end{figure}

Finally, late phase spectra provide powerful diagnostics. 
Figure 3 (left) shows the late phase nebular spectrum of SN 1998bw 
(Maeda et al. 2002, Maeda et al. 2006a). 
The spherical model results in the flat-topped 
[OI]$\lambda\lambda$6300,6363, which 
is a result of emitting O distributed in a high velocity shell. 
The jet model yields narrowly peaked [OI] if viewed from the 
$z$-direction, and doubly peaked [OI] if viewed from the 
$r$-direction. This is a consequence of the emitting 
O distributed in a disk like structure (Fig. 1). 
The jet-driven model viewed from the 
$z$-direction yields the best fit, although the other models 
fail to reproduce the characteristic [OI] profile. 

The jet model as viewed from the 
$z$-axis can explain all the observations available for SN 1998bw 
in the optical window. This is not the case for the spherical 
hypernova model (Maeda et al. 2003a, Maeda et al. 2006a, 
Maeda et al. 2006c). 
In the jet-driven model, $E_{51}$ is reduced as 
compared to the spherical model, but still $E_{51} \sim 20$ 
is much larger than canonical SNe. 
The jet-driven model for hypernovae was further strengthened 
by another example -- SN Ic 2003jd. It showed 
broad absorption lines at the early phase, 
as typical of hypernovae, although it was not 
associated with a GRB. 
SN 2003jd in the late phase showed unique 
[OI] profile - doubly peaked (Fig. 3, right panel). 
The line profile is nicely reproduced by the 
jet model viewed side ways (Mazzali et al. 2005). 
 
\section{Conclusions}
\label{sect:conclusion}

In this paper, we have presented the first study of 
deriving observational consequences of a jet-driven 
supernova model. For SN 1998bw (associated with GRB 980425), we found that 
the jet-driven model fits optical observations fairly well -- 
the model can explain most of observational facts from $\sim 1$ week since the 
explosion to $\sim 1$ year, although a spherical model can explain only 
a part of the observations. It is found that the viewing angle should be 
close to the axis of the jet, as is consistent with the fact that 
it took place with a GRB. The required kinetic energy of the SN ejecta is 
reduced to $\sim 2 \times 10^{52}$ erg as compared to the value 
suggested by the earlier spherical modeling ($\sim 5 \times 10^{52}$ erg), 
but it is still much larger than the canonical value ($\sim 10^{51}$ erg). 
This indicates that the explosion mechanism of the SNe associated with GRBs 
is totally different from that of a canonical SN.

\begin{acknowledgements}
The author would like to thank Professor Giovannelli and 
the organizers of the {\it FRASCATI WORKSHOP 2007}  
for the invitation to the extremely stimulating conference, and 
also for kind hospitality. 
The author would like to thank Masaomi Tanaka and 
Paolo A. Mazzali for participating in the {\it SAMURAI} project. 
The author is supported by the Japanese Society of the 
Promotion of Science (JSPS) under the JSPS Postdoctoral 
Fellowship for Research Abroad. 
\end{acknowledgements}

\bigskip
\bigskip
\noindent {\bf DISCUSSION}

\bigskip
\noindent {\bf QUESTION:} How much are you confident with 
the mass of a progenitor star?

\bigskip
\noindent {\bf ANSWER:} For SNe Ic, it is the ejecta mass 
that is derived by this kind of study. This is directly related 
to the mass of a C+O star just before the explosion, 
with small uncertainty in the unknown central remnant's mass. 
The derived mass of the C+O star is thus more or less reliable. 
However, we need a specific evolutionary model 
to convert the C+O core mass to the main-sequence mass. 
We assume a standard evolutionary scenario, but 
GRB progenitors may well follow a non-standard evolution. 
This is indeed an interesting subject to pursue. 

\bigskip
\noindent {\bf QUESTION:} You showed a scaling law for 
the peak date ($t_{\rm peak}$) of a SN light curve 
($t_{\rm peak} \propto M_{\rm ej}^{3/4} E_{51}^{-1/4}$). Is it sensitive 
to the amount or distribution of $^{56}$Ni? 

\bigskip
\noindent {\bf ANSWER:} The scaling law is derived by 
equating the expansion time scale and the diffusion time scale. 
It is a function of opacity, density, and the distance between 
the surface and the position of $^{56}$Ni. 
Increasing the amount of $^{56}$Ni of Fe-peak elements 
increases the opacity, thus delays the peak date. 
In the case of the GRB-associated SNe, this effect is not large, 
since the opacity is dominated by abundant O-rich materials. 
The $^{56}$Ni distribution is more important in this case. 
Indeed, the jet-driven model has the extended $^{56}$Ni 
distribution, and this is one of reasons why we can 
fit the peak data of SN 1998bw with smaller energy 
than in spherical models.

\end{document}